\documentclass[12pt]{article}  
\def\preprint#1{\thispagestyle{empty}~\newline\vspace*{-22.65mm}
\begin{flushright}\begin{tabular}{l} #1 \end{tabular}
\end{flushright}\vspace{1cm}}

\begin{document}   
\def\fecha{16 January 1997}

\markright{Circular strings, wormholes and minimum size}               
\preprint{hep-th/9608004\\ 
Phys. Rev. {\bf D55} (1997) 7872}  
\begin{center}
{\Large\bf Circular strings, wormholes, and minimum size}
\\[3mm]
{Luis J. Garay, Pedro F. Gonz\'alez-D\'{\i}az,\\[2mm]
Guillermo A. Mena Marug\'an, and Jos\'e M. Raya\\[2mm]
{\small \it Centro de F\'{\i}sica Miguel A. Catal\'an,
 Instituto de Matem\'aticas y F\'{\i}sica Fundamental,
CSIC\\
  Serrano 121, 28006 Madrid, Spain}}\\[2mm]
{\small \fecha}\\[5mm]
\end{center}

\begin{abstract}

The quantization of circular strings in an anti--de Sitter
background spacetime is performed, obtaining a discrete spectrum
for the string mass. A comparison with a four-dimensional
homogeneous and isotropic spacetime coupled to a conformal scalar
field shows that the  string radius   and the scale factor
have the same classical solutions and that the quantum theories
of these two models are formally equivalent. However, the
physically relevant observables of these two systems have
different spectra, although they are related to each other by a
specific one-to-one transformation. We finally obtain a discrete
spectrum for the spacetime size of both systems, which presents a
nonvanishing lower bound.

\vspace*{5mm}
\noindent
PACS: 11.25.Db, 04.60.Ds, 04.60.Kz\\
Keywords: strings, wormholes, algebraic quantization, minimum length
\end{abstract}
%\vspace*{5mm}

\section{Introduction}
\setcounter{equation}{0}

Circular strings in curved backgrounds have been systematically
studied \cite{ls94,ls94a,vl94,ve94},   showing interesting deviations
from their behavior in flat spacetimes. Exact classical solutions
were found in Ref. \cite{ls94}, a canonical analysis was
performed in Ref. \cite{ls94a}, and a semiclassical quantization
was carried out in Ref. \cite{vl94}.  

One can realize that circular strings in an anti--de Sitter
background classically expand and contract \cite{ls94} in the
same form as the scale factor of a four-dimensional homogeneous
and isotropic spacetime conformally coupled to a massless scalar
field in the presence of a negative cosmological constant
\cite{bg96}. In the Lorentzian regime, they both expand from zero
to a maximum radius and then recollapse, following the same
geometrical pattern. In the Euclidean regime, two asymptotically
large regions of the string world sheet are connected by a throat
just as happens in the wormhole minisuperspace model of Ref.
\cite{bg96}. One of the aims of this work is to find out to what
extent this equivalence between circular strings and wormholes
survives quantum mechanically. 

We perform a proper quantization of circular
strings in an anti--de Sitter background spacetime following the
algebraic quantization program \cite{ASH}. The same procedure was
used for the study of quantum wormholes in anti--de Sitter
spacetime in Ref. \cite{bg96}, so that the comparison between
these two quantized models can be easily carried out. We will see
that, although the quantum physics in these two systems is not
identical, an equivalence can be nevertheless recovered by means
of a unitary transformation.

In addition, as a result of the quantization process, we will
obtain the mass spectrum of circular strings, whose
semiclassical limit had already been found in Refs.
\cite{ls94a,vl94}.

Finally, we will discuss another related analogy between circular
strings and wormholes: the existence of a minimum length. That a
minimum length may appear in string theory and quantum gravity
has been suggested from various points of view (see, e.g., Ref.
\cite{ga95}). Most of them involve semiclassical reasonings. In
the models discussed here, the throat size for both strings and
wormholes can actually be quantized. We will see that the
corresponding spectra are discrete and possess a nonvanishing
lower bound.

\section{Classical solutions for circular strings}
\setcounter{equation}{0}
\label{secclass}

We will start with the string action (see, for instance, Ref.
\cite{gs88}) 
\begin{equation}
S=-\frac{1}{4\pi\alpha'}\int d^2\sigma 
\sqrt h h^{\alpha\beta}
g_{\mu\nu}\partial_\alpha X^\mu \partial_\beta X^\nu
\end{equation}
where $\sigma^\alpha=(\tau,\sigma)$ are the coordinates,
$h_{\alpha\beta}$ is the two-metric in the world sheet, $X^\mu$
$(\mu=1,\ldots, 4)$ are the coordinates on the background target
spacetime, $g_{\mu\nu}$ is its four-dimensional metric, and
$\alpha'$ is the inverse string tension.

We now reduce our model by imposing anti--de Sitter symmetry in
target spacetime and considering circular strings on the
equatorial plane. More explicitly, we will consider background
spacetimes of the form
\begin{equation}
d{\rm s}^2=s\xi(r)dt^2+\frac{1}{\xi(r)}dr^2+
r^2d\Omega_2^2,
\label{metric}
\end{equation}
where the radial coordinate $r$ must be positive, $d\Omega_2^2$
is the unit metric on the two-sphere, the function $\xi(r)$ has
the form
\begin{equation}
\xi(r)=1+\lambda r^2,
\end{equation}
$s$ takes the values $\pm 1$ depending on whether the
Euclidean ($s=+1$) or Lorentzian ($s=-1$) regime is being
considered, the cosmological constant is $-3\lambda$, and
\begin{equation}
X^1=t(\tau), 
\hspace{7mm}
X^2= r(\tau), 
\hspace{7mm}
X^3= \theta=\frac{\pi}{2},
\hspace{7mm}
X^4=\sigma.
\end{equation}
We will work in the gauge
\begin{equation}
h_{\alpha\beta}=\left(
\begin{array}{cc} sN^2 & 0 \\ 0 & 1 \end{array}
\right).
\end{equation}
The conformal gauge is achieved when the lapse function $N$ is
set equal to 1.

Then, denoting the derivative with respect to $\tau$ by an
overdot, the action takes the canonical form 
\begin{equation}
S=\int d\tau \left( p_t \dot t+p_r\dot r-N H\right),
\end{equation}
where the Hamiltonian constraint $H$, which is the only surviving
constraint (coming from the fact that the world sheet metric
$h_{\alpha\beta}$ is non dynamical), is
\begin{equation}
H=\frac{1}{2}\left( \frac{\alpha'}{\xi(r)}p_t^2+
s\alpha'\xi(r)p_r^2-\frac{r^2}{\alpha'}\right),
\label{hamilt}
\end{equation}
$p_t$ and $p_r$ being the momenta canonically conjugate to the
variables $t$ and $r$, respectively.

Let us briefly summarize the classical propagation of this system
\cite{ls94}. The variable $t$ is cyclic and therefore
\begin{equation}
\dot p_t=0,
\end{equation}
so that $p_t$ is a constant of motion. Introducing this result in
the Hamiltonian constraint, together with the expression of $p_r$
in terms of $ \dot r$, we obtain the equation (in the conformal
gauge $N=1$)
\begin{equation}
\dot r^2+s[\alpha'^2p_t^2-r^2\xi(r)]=0,
\label{rpunto}
\end{equation}
which can be solved in terms of Jacobian elliptic functions
\cite{as70}
\begin{equation}
r(\tau)=r_m \left|{\rm cn}
\left(\frac{\sqrt{-s}\tau}{ \sqrt{1-2m}},m\right)\right|.
\label{dotr}
\end{equation}
The parameters $m$ and $r_m$ are functions of the constant of
motion $p_t$:
\begin{equation}
m=\frac{1}{2}
\left(1-\frac{1}{\sqrt{1+4\alpha'^2\lambda p_t^2}}\right),
\hspace{7mm}
r_m=\sqrt{\frac{1}{\lambda}\frac{m}{1-2m}},
\label{solclas}
\end{equation}
$r_m$ being the turning point of the potential in Eq.
(\ref{rpunto}). The absolute value has been taken in order to
account for the physical restriction $r>0$. Notice then that Eq.
(\ref{dotr}) is just a formal solution at the singularity $r=0$.

The equation of motion for the variable $t$ (again in the
conformal gauge $N=1$) is 
\begin{equation}
\dot t=\alpha'\frac{p_t}{\xi(r)},
\end{equation}
and its solution is
\begin{equation}
t(\tau)=\alpha'p_t\int_0^\tau\frac{d\eta}{1+\lambda r(\eta)^2}=
\frac{2}{\sqrt{\lambda}}\sqrt{\frac{m(1-2m)}{1-m}} 
\Pi\left(\arccos \frac{r}{r_m},\frac{m}{1-m},m\right),
\label{t(r)}
\end{equation}
where $\Pi(\phi,n,m)$ is the elliptic integral of the third kind
\cite{as70} and it is understood that the function $\arccos$ 
is defined modulo $\pi$ because of the absolute value in Eq.
(\ref{dotr}).

As shown in Ref. \cite{ls94}, inserting this classical solution
in the metric element (\ref{metric}) for anti--de Sitter
spacetime, we obtain the world sheet metric 
\begin{equation}
d{\rm s}^2=r(\tau)^2(sd\tau^2+d\sigma^2).
\label{classstring}
\end{equation}

In the Lorentzian regime ($s=-1$), $r(\tau)$ is described by a
oscillating Jacobian elliptic function whose period is
\begin{equation}
T=2\sqrt{1-2m}K(m),
\label{pertau}
\end{equation}
where $K(m)$ is the complete elliptic integral of the first kind
\cite{as70}. The metric (\ref{classstring}) describes a circular
string that starts expanding from zero radius (at $\tau =-T/2 $)
until it reaches a maximum radius (at $\tau=0$) and then
recollapses again (at $\tau =+T/2 $). The same dynamics has been
found in Ref. \cite{bg96} for the scale factor of a homogeneous
and isotropic spacetime conformally coupled to a massless scalar
field in the presence of a negative cosmological constant. In the
Euclidean regime, the classical solution describes a world sheet
that is just the two-dimensional analogue of the anti--de Sitter
wormhole solutions also obtained in Ref. \cite{bg96}.

For vanishing cosmological constant, the classical equations of
motion have the solution
\begin{equation}
r(\tau)=r_m|\cos \sqrt{-s}\tau|,
\hspace{7mm}
t(\tau)=\alpha'p_t\tau.
\end{equation}
Here, $r_m^2=\alpha'^2p_t^2$. Again, we see that this
classical behavior is the same as that of the scale factor of the
minisuperspace model mentioned above (for $\lambda=0$)
\cite{bg96}.

\section{Quantization of circular strings}
\setcounter{equation}{0}
\label{secquantum}

Starting from the anti--de Sitter case, we now proceed to quantize
circular strings. The formalism for vanishing cosmological
constant can be attained by taking the limit $\lambda\rightarrow 0$.

In the Lorentzian regime, the classical variable $r(\tau)$ is
periodic with period $T$ given in Eq. (\ref{pertau}), and it then
follows from Eq. (\ref{t(r)}) that the variable $t(\tau)$
satisfies 
\begin{equation}
t(\tau+T)=t(\tau)+t(T),
\end{equation}
where
\begin{equation}
t(T)=
\sqrt{\frac{4m}{\lambda}\frac{1-2m}{1-m}} 
\Pi\left(\frac{m}{1-m},m\right)\equiv
{\cal T}(\alpha'p_t),
\label{perio}
\end{equation}
the function $\Pi(n,m)$ being the complete elliptic integral of
the third kind \cite{as70}. From this property, we can easily
check that the classical solution $r(t)$ is periodic in $t$ with
period ${\cal T}$. Consequently, $2\pi t/{\cal T}$ can be
regarded as an angular variable defined on $S^1$. Besides, the
classical solution does not depend on the sign of it. Then, time
reversal symmetry allows us to restrict $p_t$ to be positive. We
can perform a canonical change of variables from $(t,p_t)$ to
$(\beta,j)$ defined by
\begin{equation}
\beta= \frac{2\pi t}{{\cal T}(p_t)},
\hspace{1cm}
j=\frac{1}{2\pi\alpha'}\int_0^{\alpha'p_t}dz {\cal T}(z)
\equiv G(p_t^2),
\end{equation}
where, in terms of the modulus $m$ defined in Eq.
(\ref{solclas}), the function $G$ has the form\footnote{This
function coincides with the function $W(k)$ found in Ref. 
\cite{vl94}  
up to a factor $4\pi$.}
\begin{equation}
G(p_t^2)=\frac{1}{\pi\alpha'\lambda}
\frac{1}{\sqrt{1-2m}}
\left\{m\Pi\left(\frac{m}{1-m},m\right)+
\left[ (1-m)K(m)-E(m)\right]\right\},
\label{fung}
\end{equation}
$E(m)$ being the complete elliptic integral of the second kind
\cite{as70}. One can check that
\begin{eqnarray}
&G(x)\sim\frac{1}{4}\alpha' x\hspace{7mm} &\mbox{when }
\lambda x\rightarrow 0, \\
& \hspace{2mm} 
G(x)\sim\frac{1}{2\sqrt{\lambda}} \sqrt x\hspace{7mm} 
&\mbox{when } \lambda x\rightarrow +\infty. 
\end{eqnarray}
Moreover, the function $G(x)$ is continuous, strictly increasing,
and its image is the whole positive real axis, i.e.,
$G({I\!\!R}^+)={I\!\!R}^+$. Therefore, the function $G(x)$ is
invertible in ${I\!\!R}^+ $, and so  is the change of variables
from $(t,p_t)$ to $(\beta,j)$. Note that $\beta\in S^1$,
$j\in{I\!\!R}^+$, and their Poisson Bracket is $\{\beta,j\}=1$, as
should happen for action-angle variables.

Then, the first term of the Hamiltonian (\ref{hamilt}), which
contains all the dependence on $t$ and $p_t$, can be written as
\begin{equation}
\frac{\alpha'}{2\xi(r)}p_t^2=\frac{\alpha'}{2\xi(r)}
G^{-1}(j),
\end{equation}
with $G^{-1}$ being the inverse of $G$.

Let us now concentrate on the remaining part of the Hamiltonian
constraint. We can employ the following canonical variables to
describe the radial phase space:
\begin{equation}
{\tilde h}=\frac{1}{2}\xi(r)\left(\alpha'\xi(r)p_r^2+
\frac{r^2}{\alpha'}\right)
\end{equation}
and
\begin{equation}
{\tilde \theta}=
\int^r_{r_{m}}\frac{dz}{\xi(z)\sqrt{2\alpha'{\tilde h}-z^2\xi(z)}}
=\frac{(1-2m)^{3/2}}{1-m}\Pi\left(\arccos\frac{r}{r_m},
\frac{m}{1-m}, m\right),
\end{equation}
where the parameter $m$ is given in Eq. (\ref{solclas}) with
$p_t^2$ replaced by $2\tilde h/\alpha'$ and, as before, it is
understood that the function $\arccos$ is defined modulo $\pi$. 
Note that $\tilde h$ is always positive. Also, $r({\tilde
\theta},{\tilde h})$ can be shown to be periodic in ${\tilde
\theta}$ with period ${\cal T}(\sqrt{2\alpha'{\tilde
h}})/\sqrt{2\alpha'{\tilde h}}$ and one can see that this period
lies within the interval $(0,\pi)$ for positive $\tilde h$. 

Another canonical transformation provides us with the
angle-action variables for the radial part 
\begin{equation}
\theta=\frac{2\pi\sqrt{2\alpha'{\tilde h}}}{{\cal
T}(\sqrt{2\alpha'{\tilde h}})}{\tilde \theta},
\hspace{7mm}
h=\frac{1}{2\pi}\int_0^{\tilde h}dz\frac{{\cal
T}(\sqrt{2\alpha'z})}{\sqrt{2\alpha'z}}=G\left(\frac{2{\tilde
h}}{\alpha'}\right),
\end{equation}
where the function $G(x)$ is given in Eq. (\ref{fung}). As
corresponds to action-angle variables, $\theta\in S^1$ and
$h\in{I\!\!R}^+$. In terms of the angle-action variables
$(\beta,j)$ and $(\theta,h)$, the Lorentzian Hamiltonian acquires
the expression 
\begin{equation}
H= \frac{\alpha'}{2\xi(r)}\left(G^{-1}(j)-G^{-1}(h)\right),
\label{hjh}
\end{equation}
where $r$ is to be understood as a function of $\theta$ and $h$. 

We now introduce annihilation and creation variables for both
pairs of angle-action variables
\begin{eqnarray}
A_r=\sqrt h e^{-i\theta},
&\hspace{7mm}&
A^{\dag}_r=\sqrt h e^{i\theta},\nonumber\\
A_t=\sqrt j e^{-i\beta},
&\hspace{7mm}&
A^{\dag}_t=\sqrt j e^{i\beta},
\end{eqnarray}
which, as usual, verify $\{A_z,A^{\dag}_z\}=-i$ and $\bar
A_z=A^{\dag}_z$ for $z=r,t$; i.e., the set $\{A_r,\ A^{\dag}_r,\ A_t,\
A^{\dag}_t\}$ is closed under Poisson Brackets and complex
conjugation. With them, we can construct the step
variables 
\begin{equation}
J_+=\frac{1}{\sqrt 2}A^{\dag}_r A^{\dag}_t,
\hspace{7mm}
J_-=\frac{1}{\sqrt 2}A_r A_t,
\end{equation}
which, together with $j$ (or, equivalently, $h$), form an
overcomplete set of classical observables. Indeed, if we take
into account that the only solution to the Hamiltonian constraint
(\ref{hjh}) is $j=h$, it is easy to check that they weakly
commute with the Hamiltonian constraint. Again, the set 
$\{j,\ J_+,\ J_-\}$ is closed under complex conjugation,
\begin{equation}
\bar j=j\in{I\!\!R}^+,
\hspace{7mm}
\bar J_+=J_-,
\end{equation}
as well as under Poisson brackets, since, on the
constraint surface, they generate the Lie algebra of $SO(2,1)$:
\begin{equation}
\{ J_+, J_-\}=i j,
\hspace{7mm}
\{ J_+, j\}=i J_+,
\hspace{7mm}
\{ J_-, j\}=-i J_-.
\label{lieso}
\end{equation}

In order to quantize the system, we need to promote the classical
observables to quantum operators that act on a vector space. If
we introduce the variables
\begin{equation}
\tilde t=\frac{1}{\sqrt 2}(A_t+A^{\dag}_t),
\hspace{7mm}
\tilde r=\frac{1}{\sqrt 2}(A_r+A^{\dag}_r)
\end{equation}
and their canonically conjugate  momenta, we can choose as 
representation space the vector space of
complex functions on ${I\!\!R}^2$ spanned by the basis
\begin{equation}
\psi_{nm}(\tilde r,\tilde t)=\varphi_n(\tilde r)\varphi_m(\tilde t),
\hspace{7mm}
\tilde r, \tilde t\in{I\!\!R},
\end{equation}
where $n,m$ are non-negative integers and $\varphi_n(x)$ are
the normalized harmonic oscillator wave functions. The classical
annihilation and creation variables $A_z$ and $A^{\dag}_z$ are now
promoted to linear operators on this space, with action on the
basis $\psi_{nm}$  given by
\begin{eqnarray}
\hat{A}_{t}\psi_{nm}=\sqrt{m} \psi_{n(m-1)},
&\hspace{7mm}&
\hat{A}^{\dag}_{t}\psi_{nm}=\sqrt{m+1} \psi_{n(m+1)},\nonumber\\
\hat{A}_{r}\psi_{nm}=\sqrt{n} \psi_{(n-1)m},\hspace{1mm}
&\hspace{7mm}&
\hat{A}^{\dag}_{r}\psi_{nm}=\sqrt{n+1} \psi_{(n+1)m}.
\label{Apsi2}
\end{eqnarray}
We also promote the classical observables $j$, $h$, and $J_\pm$ to
quantum operators $\hat \jmath$, $\hat h$ and $\hat J_\pm$
defined as
\begin{eqnarray} 
\hat \jmath=\frac{1}{2}(\hat{A}^{\dag}_t\hat{A}_t+\hat{A}_t
\hat{A}^{\dag}_t),
&\hspace{7mm}&
\hspace{2.5mm}\hat h=\frac{1}{2}(\hat{A}^{\dag}_r\hat{A}_r+
\hat{A}_r \hat{A}^{\dag}_r),\nonumber\\
\hat{J}_+=\frac{1}{\sqrt{2}}\hat{A}^{\dag}_{t}\hat{A}^{\dag}_r,
\hspace{13.5mm}
&\hspace{7mm}&
\hat{J}_-=\frac{1}{\sqrt{2}}\hat{A}_{t}\hat{A}_r,
\label{Jaa}
\end{eqnarray}
and whose action on the basis $\psi_{nm}$ is 
\begin{eqnarray}
\hat h\psi_{nm}=(n+1/2)\psi_{nm},\hspace{1mm}
&\hspace{7mm}&
\hat J_+\psi_{nm}=\frac{1}{\sqrt 2}\sqrt{(n+1)(m+1)}
\psi_{(n+1)(m+1)},
\nonumber\\
\hat \jmath\psi_{nm}=(m+1/2)\psi_{nm},
&\hspace{7mm}&
\hat J_-\psi_{nm}=\frac{1}{\sqrt 2}\sqrt{nm}
\psi_{(n-1)(m-1)}.
\label{jhat}
\end{eqnarray}

We can choose as the inner product in our representation space
that of $L^2({I\!\!R}^2)$, which guarantees that the complex
conjugation relations can be directly translated to adjointness
relations between our operators. Since the functions $\psi_{nm}$
are orthonormal with this inner product, they provide a spectral
resolution of the identity in terms of eigenfunctions of $\hat
\jmath$ and $\hat h$. Recalling that the function $G$, and hence 
$G^{-1}$, is continuous,  the spectral theorem \cite{gp74} allows
us then to define the operators $G^{-1}(\hat \jmath)$ and 
$G^{-1}(\hat h)$ in the form
\begin{equation}
G^{-1}(\hat h)\psi_{nm}=G^{-1}(n+1/2)\psi_{nm},
\hspace{7mm}
G^{-1}(\hat \jmath)\psi_{nm}=G^{-1}(m+1/2)\psi_{nm}.
\end{equation}
Hence, the quantum solutions to the Hamiltonian constraint form the
subspace $V_p$ spanned by the wave functions $\psi_{nn}$. 
The Hilbert space of physical states is then just the Hilbert
completion $H_p$ of $V_p$ with respect to the inner product of
$L^2({I\!\!R}^2)$.

Finally, from Eq. (\ref{jhat}), it is easy to see that $\hat
\jmath$ and $\hat h$ coincide on $H_p$ and $\hat \jmath$, $\hat
J_\pm$ leave the space $H_p$ invariant. Therefore, we
conclude that $\hat \jmath$ and $\hat J_\pm$ are quantum
observables. Furthermore, under commutators, they generate the
Lie algebra of $SO(2,1)$ on the physical space $H_p$. Thus, $H_p$
carries a linear representation of the algebra of physical
observables, which, in addition, can be seen to be irreducible.

\section{The mass spectrum}
\setcounter{equation}{0}
\label{secspectrum}

The   momentum $p_{t}$  is a conserved quantity that is associated
with an isometry of anti--de Sitter spacetime. It generates
translations along the timelike Killing direction. Thus, we can
identify $p_{t}$ with the string mass $M$. Indeed, a detailed
calculation, using the embedding of the four-dimensional anti--de
Sitter spacetime in a flat five-dimensional one, shows that the
Casimir of the anti--de Sitter group is given by $C= \alpha'
p_{t}^{2}$. Since we are considering homogeneous and isotropic
configurations, the Casimir will directly give the dimensionless
string mass, i.e., $C= \alpha' M^{2}$, and, consequently,
\begin{equation}
M=p_{t}.
\label{masa}
\end{equation}
This result coincides with the semiclassical expression \cite{vl94}
\begin{equation}
M=\frac{d S_{\rm class}}{d {\cal T}},
\end{equation}
where the action of the classical solution is 
\begin{equation}
S_{\rm class}=\frac{2}{\alpha'\lambda}\frac{1}{\sqrt{1-2m}}
\left[E(m)-(1-m)K(m)\right]
\end{equation}
and $\cal T$ is given in Eq. (\ref{perio}).
From Eq. (\ref{masa}), we can obtain the    circular string 
mass spectrum
\begin{equation}
M_n^2=G^{-1}(n+1/2).
\end{equation}

As we have seen, in the limit of vanishing cosmological constant,
$G(x)\sim\alpha' x/4$, so that the mass spectrum in this limit
turns out be
\begin{equation}
\alpha'M_n^2\sim 4(n+1/2)+{O}(\lambda^{3/2}\ln\lambda),
\hspace{7mm}
\mbox{when } \lambda\rightarrow 0,
\end{equation}
which corresponds to the well-known result for flat spacetime,
except for the ground state, as the contribution from 
nonhomogeneous modes have not been considered here \cite{gs88}.

We can also expand $G^{-1}(n+1/2)$ for large values of $n$ and
fixed nonvanishing $\lambda$, obtaining the following asymptotic
behavior of the mass spectrum
\begin{equation}
\alpha'M_n^2\sim 
4\alpha'\lambda n^2
+\frac{32\sqrt{\pi\alpha'\lambda}}{\Gamma\left(\frac{1}{4}\right)^2}
n^{3/2}
+\left( 4\alpha'\lambda+
\frac{128\pi}{\Gamma\left(\frac{1}{4}\right)^4}\right)n
+{ O}(\sqrt n),
\end{equation}
whose leading term coincides with that obtained  in Ref.
\cite{vl94}.

\section{Minimum size in strings and gravity}
\setcounter{equation}{0}
\label{secminimum}

The classical solution for the radius   $r(\tau)$ of a
circular string is entirely equivalent to the classical scale
factor of a homogeneous and isotropic four-dimensional spacetime
conformally coupled to a massless scalar field, as we have
already pointed out. In view of this geometrical coincidence, we
now address the question of whether an analogous relation between
both systems still holds quantum mechanically. 

Actually, there exists an obvious isomorphism between the Hilbert
spaces and the algebras of observables of both models that
preserve the adjointness relations among the observables (see
Ref. \cite{bg96}). In this sense, we can say that these two
systems are quantum mechanically equivalent. However, we will see
that observables with direct physical meaning, such as the
maximum radius of the string and that of the universe, have
different spectra. Thus, the physical features reduce, in
principle, the above equivalence to a formal level. Despite the
physical differences between both models, they can be nontheless 
related to each other by a well-defined transformation
mediated by the function $G$, given in Eq. (\ref{fung}). This
transformation preserves besides some features of these systems,
as it happens to be the case for the existence of a 
minimum invariant size \cite{ga95}. 

Let us see in some detail how this minimum size appears. 
We will first consider the case of circular strings. We
have seen that the observable $\hat p_t^2$ has a discrete
spectrum, its eigenvalues being $G^{-1}(n+1/2)$. From this
observable $\hat p_t^2$ (and using the spectral theorem), we can
construct another observable
\begin{equation}
\hat R=\frac{1}{\sqrt{2\lambda}}
\sqrt{(1+\alpha'^2\lambda \hat p_t^2)^{1/2}-1},
\label{rmax}
\end{equation}
which also has a discrete spectrum 
\begin{equation}
R_n=\frac{1}{\sqrt{2\lambda}}\sqrt{
\left[1+\alpha'^2\lambda G^{-1}(n+1/2)\right]^{1/2}-1},
\label{rspec}
\end{equation}
which, for $\lambda=0$, reduces to $R_n=\sqrt{\alpha'(n+1/2)}$.

We see that the spectrum of the observable $\hat R$ is bounded
from below, its smallest eigenvalue being 
\begin{equation}
R_0=\frac{1}{\sqrt{2\lambda}}\sqrt{
\left[1+\alpha'^2\lambda G^{-1}(1/2)\right]^{1/2}-1},
\end{equation}
so that the mean value of $\hat R$ in any state $|\psi\rangle$ of
the string must be larger than or equal to $R_0$, i.e., $\langle
\psi |\hat R|\psi\rangle\geq R_0$. From Eqs. (\ref{solclas}) and
(\ref{rmax}), we can interpret $\hat R$ as the observable
corresponding to the radius of a circular string at the time of
maximum expansion, which coincides with the turning point of the
potential in Eq. (\ref{rpunto}). Since it is bounded from below,
we reach the conclusion that this quantity has a minimum value.
In other words, there are no quantum circular strings with a
spacetime size smaller than $R_0$. It could   be argued
that, rather than using the observable $\hat R$, we could have
employed the string coordinate radius $\hat r$, which may acquire
smaller values. However, this quantity is not invariant under
time reparametrizations, as can be easily checked. Thus the
radius $\hat r$ depends on the choice of the time parameter. On
the other hand, $\hat R$ is indeed time reparametrization
invariant; i.e., it is a true gauge-independent observable.

It   is also worth noting that this situation is
qualitatively different from what we find in standard quantum
mechanics. Indeed, the analogue of the minimum length that we have
derived in the case of strings cannot be obtained in standard
quantum mechanics because, in the latter, the position operator is
an observable. This is due to the fact that quantum mechanics,
although can be formulated in a time-reparametrization-invariant
way, has a preferred time parameter given {\it a priori} and,
consequently, we cannot require that quantum-mechanical
observables commute with the generator of time
reparametrizations.

We have also seen that homogeneous and isotropic spacetimes
filled with conformal matter are described by the same dynamics
as circular strings. Moreover, from the quantization carried out
in Ref. \cite{bg96}, we also obtain a discrete spectrum for the
observable $\hat A$, the analogue of the former $\hat R$ but with
$2\hat p_t^2$ replaced with the energy operator of the conformal
field and the string parameter $\alpha'$ with Newton's constant
$G_N$. This spectrum has the same form as in Eq.~(\ref{rspec})
with the function $G^{-1}$ replaced with the identity function
(times a factor $8/G_N$):
\begin{equation}
A_n=\frac{1}{\sqrt{2\lambda}}\sqrt{
\left[1+8G_N\lambda (n+1/2)\right]^{1/2}-1}.
\end{equation}
Again, it is bounded from below and discrete. The observable
$\hat A$ represents the radius of the universe at the time of
maximum Lorentzian expansion, i.e., what we can regard as the
spacetime size of the universe. The fact that this observable
$\hat A$ has a minimum value $A_0$ means that the smallest size
that the universe can have is not zero, but $A_0$. 
As   happened in the string case, the observable $\hat A$,
rather than the scale factor $\hat a$, gives the
time-reparametrization-invariant size of the universe.
From the Euclidean point of view, $\hat A$ represents the
wormhole throat radius, and the fact that $\langle\psi|\hat
A|\psi\rangle\geq A_0$ implies that tunneling effects in large
spacetimes mediated by wormholes will also have this minimum
size. $A_0$ can thus be regarded as the smallest wormhole throat
radius.

Furthermore, this may also have consequences for the low energy
effective physics in a background (flat or anti--de Sitter)
spacetime. In this case, it can be argued that there exists a
minimum uncertainty in the position. Indeed, the quantum
fluctuations of spacetime would have a minimum spacetime volume
of Planck's order. This would amount to have an uncertainty in
the metric of the same order and, consequently, in the
determination of any spacetime distance. On the other hand, these
spacetime fluctuations would have associated with them
fluctuations of the matter fields that would give rise to a bare
vacuum energy. It has been proposed that wormholes can
effectively drive the cosmological constant to zero as seen from
the low energy point of view \cite{co88b} even though, at the
more fundamental level, it will be present at least in the form
discussed above.

This scenario is in complete agreement with many other analyses
suggesting the existence of a minimum length \cite{ga95}. Here,
we have presented a simple example where, starting from a proper
quantum theory, this minimum length appears naturally. 

In this discussion, we have employed a conformal scalar field as
representative of the matter content. Nevertheless, one could
expect that also other fields with a discrete energy spectrum,
e.g., radiation fields, would give rise to a minimum length.
Moreover, even in the absence of matter, the fluctuations of the
gravitational field might also induce a minimum scale.

\section{Conclusion}
\label{secconclusion}

In this work, we have constructed a canonical quantum theory for
circular strings in an anti--de Sitter background. As a result, we
have obtained its    mass spectrum which, in the limit of
large quantum numbers, agrees with the result of Ref.
\cite{vl94}.

We have also seen that the classical solutions for the string
radius   are the same as the classical solutions for the
scale factor of a four-dimensional homogeneous and isotropic
spacetime conformally coupled to a massless scalar field in the
presence of a negative cosmological constant and that their
quantum theories are equivalent via a suitable isomorphism.
Despite the fact that corresponding quantities with direct
physical meaning (such as the string mass and the conformal
matter energy) have different spectra, the underlying isomorphism
allows us to introduce a specific one-to-one transformation
between these two systems.

Among the physical features that they share, it is worth
noting the existence of a minimum spacetime size which, in both
cases, is a direct consequence of the discrete spectrum of the
mass-energy operator.

\section*{Acknowledgments}

The authors want to thank C. Barcel\'o for helpful discussions.
L.J.G. was supported by funds provided by DGICYT and MEC (Spain)
under Contract Adjunct to the Project No. PB94-0107. P.G.-D.
acknowledges DGICYT for financial support under Research Projects
Nos. PB94-0107 and PB93-0139, and MEC Spanish German Joint
Action No. 161.B. G.A.M.M. has been partially suported by funds
provided by MEC and DGICYT under Research Project No. PB93-0139.

%%%%%%%%%%%%%%%%%%%%%%%%%%

\end{document}